\documentclass[conference]{IEEEtran}
\usepackage{amsmath,amssymb,amsfonts}
\usepackage{algorithmic}
\usepackage{graphicx}
\usepackage{textcomp}
\usepackage{xcolor}
\usepackage{hyperref}
\usepackage{makecell}
\usepackage{listings}
\usepackage{easyReview}

\def\BibTeX{{\rm B\kern-.05em{\sc i\kern-.025em b}\kern-.08em
    T\kern-.1667em\lower.7ex\hbox{E}\kern-.125emX}}
\usepackage[
backend=biber,
sorting=none,
citestyle=numeric
]{biblatex}
\lstdefinestyle{legalContract}{
    basicstyle=\small,
    columns=fullflexible,
    backgroundcolor=\color{gray!10},
    breakautoindent=false,
}

\begin{document}

\title{A blockchain-based pattern for confidential and pseudo-anonymous contract enforcement}

\author{\IEEEauthorblockN{Nicolas Six, Claudia Negri Ribalta, Nicolas Herbaut and Camille Salinesi
\IEEEauthorblockA{Centre de Recherche en Informatique (CRI)\\
Université Paris 1 Panthéon-Sorbonne, Paris, France\\
\{nicolas.six, claudia.negri-ribalta, nicolas.herbaut, camille.salinesi\}@univ-paris1.fr}}
}

\maketitle

\begin{abstract}
Blockchain has been praised for its capacity to hold data in a decentralized and tamper-proof way. 
It also supports the execution of code through blockchain's smart contracts, adding automation of actions to the network with high trustability.  
However, as smart contracts are visible by anybody on the network, the business data and logic may be at risk, thus companies could be reluctant to use such technology.
This paper aims to propose a pattern that allows the execution of automatable legal contract clauses, where its execution states are stored in an on-chain smart-contract and the logic needed to enforce it wraps it off-chain.
An engine completes this pattern by running a business process that corresponds to the legal contract.
We then propose a pattern-based solution based on a real-life use case: transportation of refrigerated goods. 
We argue that this pattern guarantees companies pseudonymity and data confidentiality while ensuring that an audit trail can be reconstituted through the blockchain smart-contract to identify misbehavior or errors.
This paper paves the way for a future possible implementation of the solution described, as well as its evaluation.
\end{abstract} 


\section{Introduction}

Blockchain is a registry of interconnected blocks, cryptographically linked together that contains sets of transactions \cite{nofer2017blockchain}. 
This registry is shared across peers, so-called nodes, that agree on what to write on the blockchain, based on a consensus algorithm. 
Blockchain can be seen as a distributed database, only accessible for reading and appending. 
Through those characteristics, blockchain enables tamper-proof resistance, decentralization, and security features of assets. 
First-generation blockchains were implemented to support cryptocurrencies \cite{nakamoto2019bitcoin}, but in recent years some blockchains have started to support smart contracts, which in various instances are written in Turing-complete programming languages \cite{wood2014ethereum}).   
Accordingly, smart contracts are computer programs deployed on-chain that can perform actions when specific conditions are met.
Given the tamper-resistance nature of blockchain, smart contracts provide a high level of confidence that the execution of the code will follow the intended logic, and the resulting states and data will be stored in the blockchain.

Although the proper definition of a legal contract is outside this paper scope, a legal contract has these characteristics (1) it has a defined object (2) defines the rights and duties of the parties involved, and (3) it has been agreed and consented by them.
They also must be lawful.
It has been recognized by some authors that smart contracts can indeed be deemed as a form of contract \cite{werbach2017}, yet this is a controversial topic.
In reality, the usage of smart-contracts as legally binding contracts has been restricted and presented numerous challenges such as code interpretability, contract mistakes, translation of legal concepts, and acceptance recognition \cite{giancaspro2017smart}.
In the meantime, companies are reluctant to deploy smart contracts on-chain, as it poses several risks. 
For example, data protection issues have been raised about the usage of blockchain and compliance has been deemed difficult \cite{Finck2018, filippi2018}.

In this paper, we introduce a pattern named \textit{On/off-chain smart-contract binding for confidential contract enforcement pattern} that binds an off-chain smart legal contract, a blockchain smart contract, and a business process engine to augment then enforce a legal contract through a compliant business process where important data are stored off-chain and hashes or signatures on-chain, thus ensuring the confidentiality of business data and pseudonymity of companies.
The paper is organized as follow: Section \ref{section:background} presents background knowledge, then Section \ref{section:patterndescription} introduces the pattern architecture and functioning. 
Section \ref{section:solution} describes a possible solution that uses this pattern for a real-life use-case, and Section \ref{section:discussion} analyses the pattern and present the threats of validity.
Finally, related works are presented in Section \ref{section:relatedworks}.

\section{Background}
\label{section:background}

\subsection{Legal contracts}

Smart contracts were first introduced by \cite{szabo97} in the 97, who suggested that smart contracts could replace normal, paper-based contracts whilst diminishing the human and computer cost of enforcing such documents.
Concurrently, Grigg proposed the "Ricardian Contract" as a pattern for automatizing contracts \cite{grigg2004}.
This type of contract is designed to be read by both humans and machines, should be cryptographically signed, and should be legally enforceable.
As recognized by Grigg \cite{grigg2015} the difference between these, is that Ricardian contracts hold the intentions, while the smart contract is the code, the machine-readable element of the contract.

From this perspective, legal smart contracts are smart-contracts that have some sort of legal clauses. 
\cite{clack2018} discusses the automation of legal contracts in smart contracts and their challenges.
\cite{filippi2018} created the term \textit{lex cryptographia} and applies it to smart contracts, detailing how these applications are a new type of self-governing ex-ante application of laws and rules, examining the variety of legal consequences of this new technology.
\cite{clack2018} highlights the difficulties of translating legal prose into code and what can be operationalized in a contract. 
Finally, \cite{werbach2017} reflects about the legality of smart contracts and if new doctrines are necessary. Some known initiatives to create legal smart contracts are the Accord Project and Legalese\footnote{https://accordproject.org/ and https://legalese.com/ accordingly}. 

\subsection{Patterns} 

From an architectural point of view, the term pattern appeared in the 1960s when Christopher Alexander, a well-known architect, said that the greatest architectures are made with pieces that are custom-fit to each other, to assure qualities such as aesthetics, comfort, or human needs \cite{coplien1998software}.
This term was then reused by software architects as an analogy between architecture and software architecture.
Therefore, patterns can be seen as a software abstraction that can be implemented in the architecture, or one of its components can be reused given problem \cite{coplien1998software}. 
Using patterns allows architects to create new solutions to re-using existing pieces of software.
As they have been proven effective through extensive use or analysis, they help to build high-quality software that met quality requirements, if well applied.
This is currently an emerging field of research in blockchain applications. 
In \cite{xu2018pattern}, the authors introduce 15 blockchain patterns where most of them are already implemented in blockchain applications.

\section{On/off-chain smart-contract binding for confidential contract enforcement pattern}
\label{section:patterndescription}

In this sub-section, we express the pattern by using the \textit{Alexandrian form} format, based on building architecture to describe repeatable patterns \cite{coplien1998software}. 
The pattern is detailed in the \textit{Solution} part of this description.

\textbf{Context} - During past years, the different industries have been impacted by new and disruptive technologies. 
One of such technologies is smart contracts.
 One objective of the smart contracts is to translate legal contracts as code or help the legal contract by augmenting some of its clauses into machine-readable functions. 
The emergence of blockchain technologies has led to the creation of blockchain-based smart contracts, that has benefited from blockchain characteristics. 

\textbf{Problem} - When stored on-chain, data becomes tamper-resistant. 
Additionally, the utilization of smart contracts isn't always possible for stakeholders due to concerns about the irreversibility of smart contracts in certain blockchain technologies \cite{atzei}.
For example, creating GDPR compliant blockchains for storing personal data systems is still an open topic\cite{Finck2018}. Although not all smart contracts might deal with personal data, there are concerns over the leakage of organizations' confidential data or their identities, even if data are encrypted.    
With those risks in mind, we investigate how can we automate the enforcement of specific data-driven clauses of legal contracts, in a machine-trustable way, while guaranteeing the pseudonymity of parties and business data confidentiality.

\begin{figure}
    \centering
    \includegraphics[width=0.7\linewidth]{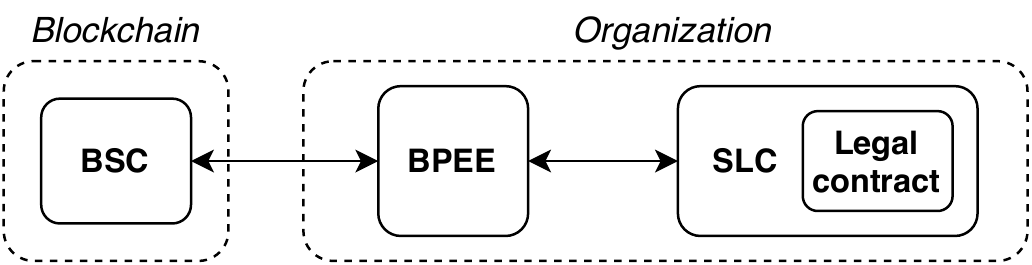}
    \caption{Pattern scheme.}
    \label{fig:scheme}
\end{figure}

\textbf{Solution} - We propose a pattern constituted of three components: a blockchain smart contract (BSC), an off-chain legal contract augmented with functions and data called Smart Legal Contract (SLC), and a business process execution engine (BPEE).
Figure \ref{fig:scheme} shows the organization of the components in this pattern.

The first component of this architecture is the SLC, which is the backbone of our system, as it contains elements related to the legal contract we want to augment.
First, it contains the legal contract, written in plain text. 
The legal contract must identify the parties, using a cryptographic mechanism that keeps their identities confidential on-chain and can only be deduced while having access to the legal contract. 
In this article, we propose the usage of a PKI, however other methods can be envisioned. 
In this pattern we define three types of parties that can be added to the contract:
\begin{itemize}
    \item Organizations - an active participant for the business process, that has interests in it. 
    \item Oracles - in blockchain, an oracle is a system entitled to bring information on-chain, as smart contracts cannot actively retrieve information \cite{xu2018pattern}. 
    Here, oracles will provide data to serve contract enforcement. 
    \item Mediator authorities - a participant that only intervenes in the business process if asked by another organization involved in the SLC. 
    Its role is mediating between organizations and enforce exceptional clauses.
\end{itemize}
Note that passive participants, such as devices (oracles) may not need the SLC as their only purpose is to send messages that impact the contract enforcement.
Second, it embeds a data model and a data store to make available the data of the contract in an understandable format (such as names) as well as its states.
Third, it contains some logic functions to handle contract data and states, thus making it possible to enforce the clauses.
The SLC is connected to the BPEE, that drives the business process associated with the legal contract and thus, enforces the SLC clauses if needed.
The requests sent to the SLC are stored.
It enables the possibility to display a complete history of requests sent to the SLC and by extension, the audit trail of the LC.
An SLC might be unique: a participant is in charge of listening to other requests, then send them to the SLC.
It might also be multiple: every participant owns its instance, and share the message and the result obtained whenever they perform a request on the SLC.
Such organizations do not prevent malicious users to enforce clauses in an abusive manner.
The BSC is the component that prevents this issue.

Compared to the other components, the BSC is unique and common to the legal contract's participants
The state of the contract (defined below) the different clauses and related agreements are stored inside the BSC. 
As the legal contract is required to interpret them, there is no leak of external information.
Only the states are at risk, but several techniques can be employed to mitigate this, such as obfuscation.
Finally, the hashes of data messages shared between participants are stored in the BSC.
A direct reference is established between the SLC and the BSC: the legal contract contains a reference to the BSC, and the BSC contains a hash of the SLC where the legal contract is signed by participants.
Thus, they recognize the role of the BSC in the business process.

The third proposed component is the BPEE.
The BPEE is the orchestrator of this architecture, as it executes a local business process derived from the global business process that describes the legal contract flow.
This means that the local business process will vary depending on the participant activity, but it may comply with the global business process by being aware of the collaborative steps of the legal contract (agreements, exceptions, etc).
The BPEE is also responsible for listening to other participants' requests and redirects them to the SLC if needed.
To be taken into account, a request must have seen its hashes stored on the BSC.
It proves its existence and will serve as proof if a participant tries to act maliciously by claiming they did not send a message, although they did.

\begin{figure}
    \centering
    \includegraphics[width=0.69\linewidth]{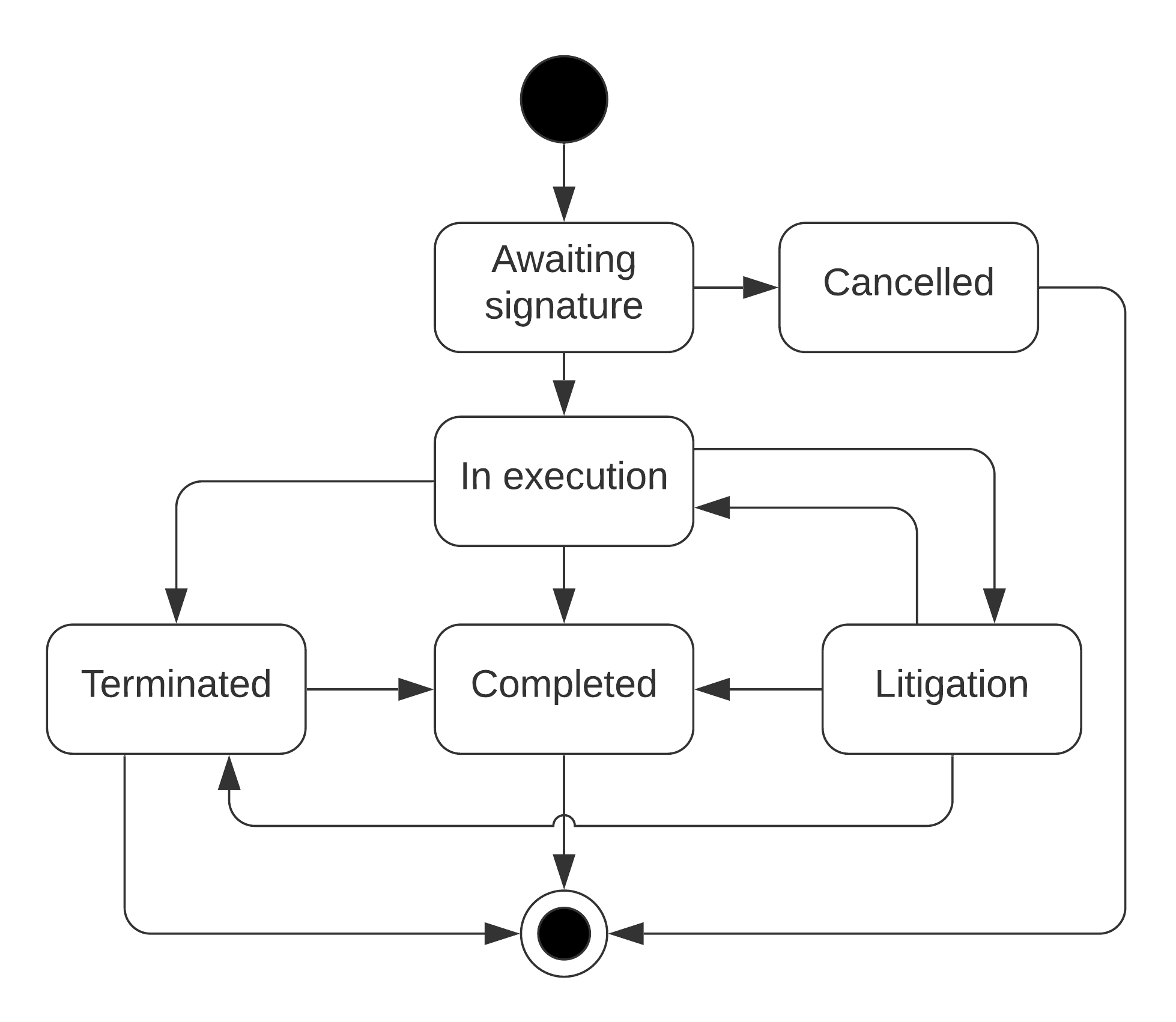}
    \caption{State transition of the BSC.}
    \label{fig:statediagram}
\end{figure}

The state of the legal contract, and by extension the state of the BSC and the SLC, may be defined as follows: \textit{Awaiting signature}, \textit{In execution}, \textit{Completed}, \textit{Litigation}, and \textit{Terminated}.
Figure \ref{fig:statediagram} shows the generic state transition during the life cycle of the contract. 
Depending of the needs of pattern users, this diagram may be modified.

From a behavioral point of view, we can describe the pattern in four parts.
First, the contract signature. 
The participants must agree on the SLC content (the legal contract and the wrapped logic) and sign it as an approval.
Whenever an SLC is fully signed, a hash of it is made and stored on-chain (BSC).
Thus, the BSC has to be deployed before this signature round.

Second, the contract execution. 
Following its signature, participants will execute their local business process that complies with the legal contract.
Thus, they may have to halt this execution to wait for steps that must be completed by other participants, such as clause enforcement or agreements on operations performed to fulfill the contract.
To perform such actions, participants have to collaborate with others by sending messages. 
To be considered valid by other participants, a message has to be stored on-chain. 
This requires a particular process to do so: 1) a hash of the message is generated, 2) this hash is stored on-chain, 3) the reference to this hash is added to the message, 4) the message is dispatched to other participants. 
Also, as multiple SLC instances of the same legal contract may exist, a request made to the SLC and the response received must be shared with others, following the same process.
This ensures that all the participants possess the latest version of the SLC at any time.
In any case, a participant can request a version of the SLC to another, and verify its authenticity using the hashes stored inside the BSC.

Third, the contract monitoring. 
At any time, a participant that possesses an instance of SLC can completely rebuild the audit trail of the legal contract. 
This includes all the enforced clauses, all the agreements made between parties, and all the important messages shared.
This also allows the quick identification of malpractices from a participant towards the good execution of the contract.

Fourth, the contract litigation.
Participants may disagree on operations performed, or maliciously alter the SLC by enforcing clauses they are not allowed to. 
In this context, the execution of the contract may be halted, and an authority able to litigate may intervene. 
This authority can be defined at contract creation or voted by other participants.
If needed, a litigation authority can delete incorrect hashes and agreements on the BSC. 
This procedure has to be implemented carefully, as this could cause inconsistencies between a current business process execution and the state of the contract.  

\textbf{Forces} - This pattern has to balance the following forces:
\begin{itemize}
    \item pseudonymity: legal contract participants must remain anonymous on-chain.
    \item Traceability: a participant can retrieve an accurate audit trail that can be done at any moment.
    \item Confidentiality: sensitive data from the execution of the legal contract business process must remain confidential.
\end{itemize}

\textbf{Example} - We present a possible solution based on this pattern in Section \ref{section:solution}.

\textbf{Force resolution} - Participants pseudonymity might be broke if a malicious participant uses a previous instance of SLC to build a list of related BSC on-chain and determine the participants' identities through the behavior of its states.
Also, this pattern requires a strong communication mechanism and data transfer protocol between participants, as they have to send data off-chain (after storing a hash of them on-chain).

\textbf{Design rationale} - Many applications and studies take profit of the blockchain smart contract as a state holder for external activities.
Regarding the binding of an off-chain and an on-chain contract, we found only one study that introduces this concept \cite{xu2018pattern}. 
However, this pattern does not ensure parties' pseudonymity or data confidentiality in its presented form.

\section{Pattern-based example solution}
\label{section:solution}

In this section, we describe a possible solution using this pattern for a problem given in Section \ref{section:patterndescription}, the transportation of refrigerated goods.

Transportation of goods can appear as a simplistic activity, but it hides many aspects of complexity, such as planning in-between organization, damages to good, transportation times or problems during the transportation time. 
These issues might cause trust problems between companies.
In this context, blockchain can help to improve trust between parties, by ensuring the traceability of the product through the supply chain. 
New research has proposed blockchain-based solutions for the transportation of refrigerated goods, such as \cite{kamath2018food}. 
This scenario will present a solution to address this problem, by leveraging the pattern presented in this paper. 
A buyer wants to buy a refrigerated good from a supplier, that needs to stay under a specific temperature. 
For simplicity, the supplier will provide the product as well as delivering it. 

At first, a buyer will send a request for proposal (RFP) to the supplier.
If the supplier accepts the request, he will provide an offer to the buyer, that contains a price grid for requested products. 
If the buyer accepts the offer, a contract is enacted between the two parties. 
After the signature of the contract, the supplier will prepare, transport, and deliver the order to the buyer. 
In the meantime, the buyer will periodically check the status of the shipment to be sure that the temperature stays under the defined threshold in the contract.
Thus, if the temperature is too high, the buyer is entitled to terminate the contract at no cost and refuse the order. 
He can also check if the order is late: if so, the buyer can claim to the seller the payment of additional penalty fees. 

\subsection{Solution structure}

This solution contains three services from our components defined in the pattern description, but also two additional services: a blockchain middleware, to make the bridge between the blockchain and the BPEE, a database to store traceability information, and a front-end application to allow users using the application.
We proceed to explain the design of such solution. 

\subsubsection{SLC component}
We designed the SLC as proposed by the Accord Project
thus in three parts.
First, the legal contract template: written in plain text, the legal contract template contains placeholders for data that varies over contracts, such as organizations' information and contract conditions.
Second, the data model, that describes the data structures held by the SLC.
Third, a collection of logic functions to enforce specific clauses. 
For this pattern, logic functions will consist of verifying and storing organizations' signatures into the SLC state, and exceptions triggered during the execution of the business process. 
The SLC also manages two types of data: static data, that is the contract information that fills its placeholders and serves the clauses, and dynamic data, that is the state of the SLC and variables such as delivery and agreement date and time.

Four participants constitute this contract: the buyer, the seller, an IoT device placed into the shipment to monitor the temperature, and a mediator authority that can litigate on conflicts.
In this solution, we use a PKI to authenticate a party on-chain and allows them to sign messages.
Many clauses of the legal contract can be enforced, from the shipment delivery to the agreement of its content by the buyer. 
To do that, the BPEE will send a request to the SLC. 
The BPEE is in charge of storing the request and the response hashes to the BSC if it is the requestor, or looking up to the BSC to check if a received message from another participant has been stored. 
Therefore, the SLC will simply take the request and enforce the clauses if its state allows it and return a response.
Every request made to the SLC and their according responses are stored in the database and forms the audit trail, attested by the BSC.
A user can, at any moment, display it using the front-end application.

\subsubsection{BPEE service}
In this solution, the BPEE runs the local business process from a participant, which must be compliant to the SLC; i.e. the business process should act accordingly at execution to the clauses enforced.
The BPEE is in charge of listening to other participant's messages and verify their traceability through the BSC.

\subsubsection{Blockchain smart contract and middleware service}

The BSC stores the data necessary to the execution of the business process on the participants' side in a machine trustable manner. 
The BSC contains a datastore of clauses. 
A clause is defined by a reference to the legal contract and contains its state (Enforced, Awaiting, Cancelled) as well as a datastore of hashes of messages sent from one participant to others.
A service, named blockchain middleware, is in charge of making the bridge between the BPEE and the BSC. 
On one side, it listens to the BPEE requests and executes them on-chain by leveraging a PKI that represents the organization. 
On the other side, it monitors events on the blockchain that affects the user (BSC creation, clause enforcement, etc). 

\subsection{Solution behavior at runtime}

\subsubsection{Contract signature}
\begin{figure}
    \centering
    \includegraphics[width=0.7\linewidth]{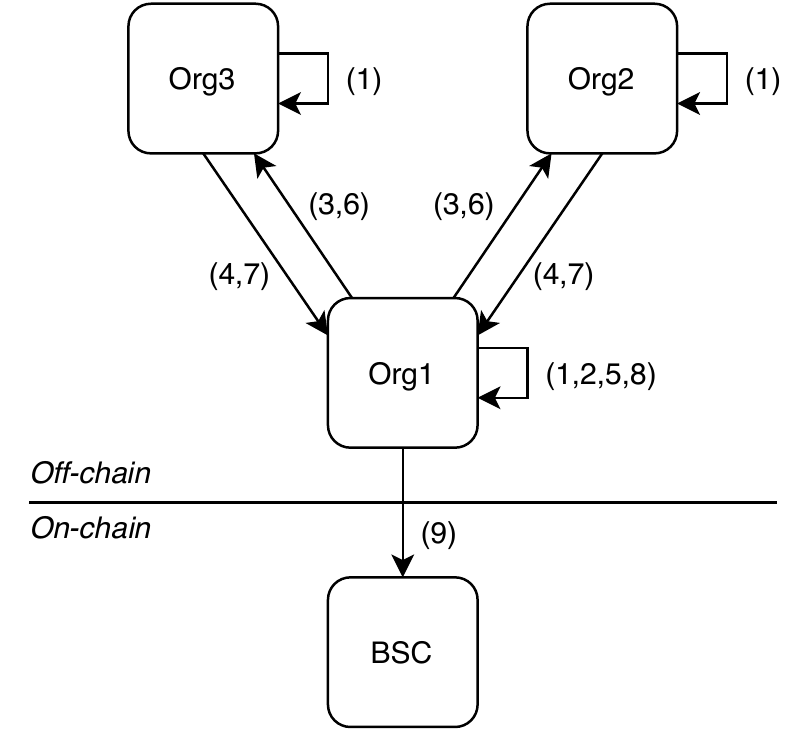}
    \caption{Contract creation and signature.}
    \label{fig:contract}
\end{figure}

Figure \ref{fig:contract} represents the steps (labeled with numbers) that need to be carried out before the execution of the contract.

To begin, each organization will create a set of public/private keys for himself and trusted oracles (1). 
Those keys are unique for every business process instance and will act as an identifier.
This ensures that no information about transaction volume or organization identity is leaked on-chain. 
Then, an organization will take the lead on writing the legal contract, modeling the business process (into a BPMN file) that respects the SLC clauses, developing the logic and data model, and collecting signatures (2).
After that, the responsible organization will request the public keys of the other organizations (3,4).
The public key will be bind to their identity into the legal contract. 
Thus, every signature made using their private key represents an agreement from the corresponding company. 
Owning the public keys, the responsible organization can generate a complete SLC instance (5) and deploy the BSC on-chain by providing the addresses of every participant, oracles, and mediators (6). 
As this is an on-chain contract, a public address for it is generated, 
returned to the responsible participant and added into the SLC (7). 
After this step, the responsible organization will send a package containing the SLC and the BSC address to other participants (8). 
If they agree about its content, they will be asked to generate a signature from the SLC data then return it (9). 
Once all signatures are collected, the responsible organization includes them into the SLC, makes a hash of the signed SLC (10), and sends it as a transaction to the BSC (11). 
At this step, the contract is signed by all the organizations, and the authenticity of their signature is now proved by the BSC. 

\subsubsection{Contract execution and enforcement}

Each participant is responsible for its actions to perform in the context of the legal contract. 
They'll use the BPEE and an according business process to pilot their activities.
Their processes must be compliant with the SLC. 
Thus, multiple activities on their business processes are linked to agreements (embed into clauses) that must be made between the buyer and the seller, as well as clauses that could be enforced and change the contract state. 
To enforce a clause, a particular suite of actions takes place. 
First, the participant must send its request to the SLC.
It will get a response indicating that a clause has been enforced or an error. 
If it gets a response, it will then hash it, as well as the request and send a message to the BSC with both hashes and the according contract state change if needed. 
The state change included in the message must tie in with the state change requested by the SLC after the enforcement of the clause off-chain.
After, the participant is finally entitled to diffuse its message to others.
They will perform the request on their contract, obtain a response, hash them, and compare the hashes to the ones in the BSC.
If they correspond, they will simply accept the message.
Otherwise, this could trigger a conflict between two or more participants.

\subsubsection{Contract mediation and monitoring}

Given the fact that participants own a copy of the messages sent throughout the execution of the contract and have access to the BSC hashes, they can, at any moment, extract the audit trail of actions performed. 
This allows highlighting of parties' misbehaving or mistakes in the execution of the contract.
When such a problem occurs (or if a participant asks to), a mediator can intervene and alter the audit trail to remove hashes, enforce clauses manually or change the state of the contract (eg. termination).
This feature must be used carefully, as it could lead to mistakes in the business process execution. 

\section{Discussion}
\label{section:discussion}
\subsection{Pattern analysis}

\subsubsection{The pattern guarantees the pseudonymity of the participants, as well as their business information} We propose the usage of a cryptography mechanism to authenticate users on-chain and off-chain. 
We used a pair of keys that are unique through instances of business process execution.
As the binding between public keys and companies' identities are kept private to the group of participants and mediators for a business process instance, it is very difficult to re-identify a company knowing only its public address.
This is also not possible to use a public address and its related identity to get information about more than one business process instance. 
Regarding the business information, all of the logic and knowledge are deported into off-chain functions and clauses. 
Thus, the BSC only contains hashes or obfuscated states. The hash protocol used should be preimage and collision resistance, for security reasons. 
This aims at ensuring pseudonymity of participants and their business information confidentiality.
Indeed, information stored into the BSC is usable only by owning and instance of the corresponding SLC. 

\subsubsection{The pattern ensures that an immutable audit trail can be retrieved} As participants store messages into a local database, and the BSC holds state changes and hashes of messages shared, it is possible at any moment to retrieve the audit trail by combining the data from those two sources. 

\subsubsection{The proofs stored on-chain following this pattern could aid or work as evidence in dispute resolutions} Given the tamper-proof nature of blockchain, the BSC \& SLC audit trail can help in dispute litigation as evidence. 
 Given that the parties have consented and entered freely in the agreement, they knew about was happening. 
 It remains to be discussed what is the level of confidence this information, given that the underlying assumption is that the data in the blockchain has been entered correctly and not tampered before. This is briefly commented in the next section.
 
\subsection{Threats of validity}

The usages of this pattern may rely on oracles, that could act maliciously for the profit of one or several participants. 
Mediators could also act maliciously, by enforcing clauses without valid reasons. 
Thus, the legal contract must take into account the possibilities of misbehaving to ensure that every participant is protected against that.
Also, policies in private blockchain might help define the behavior. 
We assume that for the system to work, off-chain data must be kept securely, as the audit trail relies on off-chain data.
Losing such data makes it impossible to consult the audit trail, thus the traceability of the business process execution is lost.

Regarding the pseudonymity and confidentiality, it may be possible to scrap every instance of BSCs on the blockchain, knowing its original bytecode. 
By inferring the clauses information and the number of parties inside a BSC to an SLC instance, it may be possible for a previous participant to identify instances of BSCs for a given SLC, thus getting knowledge about transaction volume and business process states from other participants.
This can be mitigated by using a private blockchain where a BSC instance is only accessible to instance parties, or when there are so many instances deployed that it becomes impossible to infer BSC instances and SLC information with certainty.

Finally, as recognized by \cite{Wust18}, an organization might tamper their data oracle which might inject fake data into the blockchain. This might be mitigated by the use of a private blockchain, or by defining policies of behaviour.

\section{Related works}
\label{section:relatedworks}

There are a few papers that discusses about blockchain-based patterns. 
\cite{xu2018pattern} introduces a collection of blockchain patterns for software architecture, classified by their purpose. 
One of the patterns is based on the binding between a legal and a smart contract, a similar approach to ours. 
This allows the friction-less enforcement of clauses on-chain, but this does not guarantee any anonymity or confidentiality.
Storing hashes of off-chain data or states on-chain is also an approach commonly used by academics and practitioners. 
Such a pattern exists in \cite{xu2018pattern}, and some applications introduce this system to guarantee the confidentiality of data.
In \cite{haarmann2019executing}, authors implemented a system to allow companies to execute decision models privately by only storing a hash of the decision off-chain, and revealing inputs and logic if a conflict occurs to litigate. 
Also, \cite{harer2018decentralized} proposes a system where collaborative business processes can be designed, the activities being local private processes from collaborators. 
States of those workflows are stored on-chain, to prove at any moment that parties have used a workflow compliant to the main business process.
Our pattern is using such features to fulfill its purpose while according particular attention for data confidentiality. 

\section{Conclusion}
\label{section:conclusion}

Pseudonymity and confidentiality are often \textit{sine qua non} conditions for companies to operate on blockchain, as trust is for leveraging shared business processes. 
Unfortunately, store business data and logic on-chain is an at-risk practice, even if data are encrypted.
Also, the usage of smart-contracts as legal contracts raises issues: lack of expressiveness, contract errors, or translation difficulties. 
In this paper, we introduced a pattern that binds a smart legal contract with a blockchain smart-contract to help with both of these requirements.
Then, we proposed a solution based on this pattern that addresses a real-life scenario issue: transportation of refrigerated goods.
Finally, we perform an analysis of produced work to ensures that it correctly fulfills the goals of this pattern: company pseudonymity, data confidentiality and traceability, and legal enforceability of the legal contract clauses.
Future works will aim to provide an implementation to this pattern, that conforms to the solution presented here. 
Such implementation is already in progress and available on Github\footnote{https://github.com/nicoSix/slc-implementation}.

\printbibliography

\end{document}